# A proposed superluminal S-field mediating quantum entanglement

Vitalij Garber

George Mason University, Virginia.
February 22, 2007

Abstract

We propose a bimetric space-time consisting of two vierbein bundles, in which a superluminal S-field mediates a causal exchange of quantum information associated with quantum entanglement. The resulting theory leaves the usual quantum field interactions, as well as the speed of light unchanged, but introduces quantum information effects related to the S-field vierbein bundle. We show that such S-field interactions with the Dirac field, affect its spin density, as we would expect for a field that would impart entanglement related information.

## 1. Introduction

In relativistic quantum field theory, all fields are assumed to obey the Special and General theories of relativity and are formulated in a "covariant" form. This usually means that our laws of nature must transform in such a way as to have the same form when observed from different coordinate systems. The usual metrics used to define space-time, imply that no material field can propagate faster than the speed of light. From this point of view, known interactions among the various fields, take place by means of particles (usually exchanges of various mesons) or electromagnetic quanta and are limited to the speed of light. However, quantum field theory implies *instantaneous* information transfer among the entangled quantum states, or what Einstein called "spooky action at a distance." This contradicts the constraints of the Special and General theories of relativity, and to date has not been explained (1), (2). Many physicists view this as simply a natural feature of quantum theory, unencumbered with classical causal notions of space-time. But others like Richard Feynman, call this phenomenon "*The central mystery of quantum mechanics*" (3).

John Bell showed that joint probabilities for combinations of spacelike separated events depend on the choice of particular measurement operations that are applied to entangled systems at those separate events, and moreover that this dependence is non-linear. He derived a set of inequalities which are used to assess entanglement experimentally (4). In 1981, with the advent of new technology, Alain Aspect and his colleagues, experimentally confirmed mutual effects among entangled particles for space-like separations (5). They showed "six-sigma" violation of Bell's inequalities with source-polarizer separations of up to 6.5 kilometers. This lead Alain Aspect to conclude: "*John Bell demonstrated that there is no way to understand entanglement in the framework of the usual ideas of a physical reality localized in space-time and obeying causality.*"

These advances led to a renewed interest in understanding and modeling the apparently superluminal speed of quantum information and Lorentz symmetry breaking in general. Garisto has recently provided a summary classification of the various approaches (6). We propose an approach based on the introduction of a superluminal field, which would mediate entanglement.

Many physicists are reluctant to consider a new field, especially a superluminal one, which might mediate quantum entanglement, because of perceived problems with causality. Yet there are ways to formulate a field theory to deal with such a field so as not to violate causality. We think that modeling such a field explicitly within the framework of quantum field theory, could provide not only a model for the phenomena of quantum entanglement, but indications how one might construct experiments leading to new insights. One could adopt one of several approaches to modeling faster then light or superluminal transfer of information. For example, one could use the "universal wave function" approach of deBroglie (7), as extended by Bohm and Hiley (8). Bohm pointed out that one can always introduce such an instantaneously acting field; to quote, *"one could suppose that in addition to the known type of fields there was a new kind of field which would determine a space-like surface along which non-local effects would be propagated instantaneously."* Alternatively, to avoid introducing a special coordinate system, one can introduce two separate vierbein bundles, as proposed by I. T. Drummond in his "variable light cone theories" (9 ), or the bimetric approach first proposed by N. Rosen (10 ) and recently expanded to provide a causal description of quantum entanglement by J. D. Moffat (11).

We will adopt the bimetric ideas in vierbein representation, but use a different way to introduce such a field into the dual metric "space-time." Other authors used the bimetric approach to provide for a way to introduce a causal, superluminal propagation of either *light signals* (12), or *gravitational waves* (9), (13). When addressing a causal description of quantum entanglement, Moffat assumed a structure which provided for *light* signals to communicate quantum information. In the Special Relativistic case, it was assumed that it is the speed of *light* or the *electromagnetic field* that becomes superluminal relative to the usual Lorentz space-time, when quantum interactions representing entanglement are introduced (11). When taking gravity into account, the formulation defined a new gravitational metric composed of the usual one used in General Relativity and an additional "quantum mechanical metric" representing the degree of entanglement (14).

We posit that it is *not* the gravitational or the electromagnetic fields which mediate quantum entanglement, but a new, not previously formulated, superluminal "S-field." In our bimetric approach, we assume that a vierbein bundle defines the usual Einstein gravitational metric (or in the Special Relativistic formulation, the Lorentz metric) as is the case for all current fields, but introduce an additional vierbein bundle associated with the metric of the superluminal S-field.

Thus the usual material fields will continue to be constrained by the gravitational metric chronology of the usual curved space-time and the speed of light. However, the second metric applies to a different causal structure which reflects the superluminal propagation of the S-field. Such an approach is closer to the accepted quantum field theoretical framework which assumes a constant and limited speed of *light*, and focuses on the information transfer aspects of the new field without changing the "material" interactions of our current theories.

We derive an affine structure by requiring general covariance for our equations in the introduced bimetric space-time. We will see that covariant derivatives, along with the bimetric vierbein formulation, lead to a pseudo-Riemannian space time which is no longer automatically symmetric. This result is analogous to the Einstein-Cartan theory, which is the appropriate theory of General Relativity when matter with spin is present.

In exploring matter and S-field interactions, we will start with the simplest possible model, assuming the S-field to be a *massless, scalar field*. In addition to simplicity, there are also arguments to assume the gravitational as well as the S-field variables to be *classical, c-number* quantities (15).

## 2. A bimetric space-time incorporating causal, superluminal, S-field interactions

Moffat takes the position that a "*Minkowskian metric with one light cone, is not adequate to explain the physics of quantum entanglement. The standard classical description of space-time must be extended when quantum mechanical systems are measured*" (11).

Other authors have pointed out that causality is not violated by superluminal interactions, but rather by using the chronology associated with the usual metrics of Special or General Relativity theories (16), (17). Furthermore, Bruneton (18) has analyzed this issue and points out that in a bimetric theory, the weak equivalence principle is satisfied as long as all mater fields are coupled to the same metric.

One can introduce two vierbein bundles each associated with a distinct metric. "Material" fields would propagate relative to the space-time metric defined in terms of the vierbein, equivalent to that of General Relativity:

$$\hat{h}_i^\mu(x)\hat{h}_j^\nu(x)\eta^{ij} \equiv \hat{g}^{\mu\nu}(x) \tag{1}$$

where $\eta^{ij}$ is the usual locally defined, Lorentz metric tensor. The S-field would propagate relative to its respective metric, defined by its vierbein bundle as:

$$\underline{h}_i^\mu(x)\,\underline{h}_j^\nu(x)\,\eta^{ij} \equiv \underline{g}^{\mu\nu}(x) \tag{2}$$

The full bimetric tensor, governing the interaction of the S-field, with other fields, becomes:

$$\hat{h}_i^\mu(x)\hat{h}_j^\nu(x)\eta^{ij} + \underline{h}_i^\mu(x)\,\underline{h}_j^\nu(x)\,\eta^{ij} \equiv h_i^\mu(x)h_j^\nu(x)\,\eta^{ij} \equiv g^{\mu\nu}(x), \text{ where } g^{\mu\sigma}g_{\nu\sigma} = \delta^\mu_{\ \nu} \tag{3}$$

The composite bimetric vierbein $h_i^\mu(x)$, have the usual vierbein structure, since the two vierbein bundles are related at a point x, by a local linear transformation (9):

$$\hat{h}_i^\mu(x) = \Lambda_i^j \underline{h}_j^\mu(x) \tag{4}$$

The above definitions are based on the fact that for given a vierbein field, there is a unique metric tensor field, while a metric field can not be expressed in terms of vierbein, unless it defines a Minkowskian inner product.

We can note that the resulting metric and both of its parts, are locally invariant under Lorentz transformations:

$$ds^2_1 = [(\hat{h}_i^\mu(x)\hat{h}_j^\nu(x)\eta^{ij} + \underline{h}_i^\mu(x)\,\underline{h}_j^\nu(x)\eta^{ij})]\,dx_\mu dx_\nu \tag{5}$$

In the usual sense the bimetric vierbein components can be used to go from quantities given with respect to a local bimetric coordinate system to the same quantities given with respect to a "global" bimetric coordinate system $A^\mu(x)$, and vice versa; i.e.

$A^\mu(x) = h_k{}^\mu(x) A^k(x)$ (6)
$A^k(x) = h^k{}_\nu(x) A^\nu(x)$.

This "interaction space-time" bimetric can be described by two light cones where the light cone for the S-field, can fill the upper hemisphere in the usual representation and still provide a causal formulation, even for *instantaneous* propagation (6), (18), for an angle θ, from the light cones vertical axis, up to θ= π/2 .

This approach has analogues to that of I. T. Drummond, who structured a "variable light cone theory" to deal with possible Lorentz symmetry breaking in the early universe (9). We assume that only non-degenerate values of $g^{\mu\sigma}$ correspond to meaningful space-time, or

$det(g^{\mu\sigma}) \neq 0$, (7)

where following Drummond, we further assume that the volume elements of the two vierbein bundles are the same, implying:

$det\{\hat{h}_j{}^\mu(x)\} = det\{\underline{h}_i{}^\mu(x)\} \equiv h \neq 0$. (8)

The vierbein and the affine connection variables will be considered as "physical" field variables, and the dynamics for the various field variables will be derived from a Palatini type of variation of the total action. The trajectories for interacting field variables will follow geodesics in this bimetric space-time.

Nature seems often to follow the simplest possible representations. Thus, to explore the matter and S-field interactions, we will start by assuming the S-field, $\varphi$, to be a *classical, massless, scalar* field.

While we assumed the bimetric of the gravitational and S-fields to be made up of classical vierbein variables in the quantum field theory context, there are arguments as to why such an assumption may not be only simplifying, but appears to reflect the unique nature of these fields. One can show that their respective vierbein and affine connection variables arise as c-number fields under "Lorentz gauge transformations" (19), (20) , if one maintains the usual notions of quantum measurements and the equivalence principle (15).

One can assume the scalar field to be dimensionless with the usual derivative coupling, and following Clayton, Moffat and others (13), (18), we can take it this coupling to be that part of the bimetric tensor which represents the S-field in Equation 2:

$\underline{h}_{i\mu}(x)\,\underline{h}_{j\nu}(x)\,\eta^{ij} \equiv \lambda^2 \partial_\mu\varphi \partial_\nu\varphi,$    where $\lambda$ is a coupling constant. (9)

The bimetric for the interacting fields in Equation 3 becomes:

$ds^2_I = [(\hat{h}_i{}^\mu(x)\hat{h}_j{}^\nu(x)\,\eta^{ij} + \lambda^2 \partial^\mu\varphi \partial^\nu\varphi)]\, dx_\mu dx_\nu$ (10)

## 3. The S-field interacting with matter

We are now in a position to construct the action for the S-field interaction with the matter fields.

In addition to the vierbein $h_l^\nu(x)$, we will introduce the bimetric analogues of affine connection field variables $A^{kl}{}_\mu(x)$, and treat them as independent c-number fields interacting with matter field variables that are q-number operator functions. Thus the total action, composed of the bimetric interaction field (analogous to the gravitational field in the usual formulation of quantum field theory, but now composed of both gravitational and S-field vierbein), and the action associated with the matter (where we specifically choose the Dirac field), can be written as:

$$A = \int L_{IB}\left[h_k^\nu(x), A^{lk}{}_\mu(x)\right] h d^4x + \int \langle \, |L_D\left[h_k^\mu(x), A_\mu^{kl}(x), \overline{\psi}(x), \partial_\mu \psi(x)\right] \, \rangle h d^4x \qquad (11)$$

The Dirac field $\psi$, was chosen for illustration, because it's spin one-half nature is intimately related to the entanglement phenomena, though similar results can be obtained for the electromagnetic field. The expectation value of the last integral is in the state of the system when measurement occurs. The invariant volume element $h d^4x$ is formed from the determinant of the vierbein matrix, i.e. $h(x) = \det h_l^\nu(x)$. As usual, the Euler-Lagrange equations of motion will follow from the variation of the action.

Similarly to the usual affine connection formulations of the Dirac field interacting with gravity, notions of general covariance can be applied to the bimetric framework. For transformations of the coordinate systems from the global point of view, one needs to replace ordinary derivatives by covariant derivatives :

$$\partial'_\mu \psi'(x') \to [\partial'_\mu + S(x') \partial'_\mu S^{-1}(x')] \psi(x'), \qquad (12)$$

This arises from the "Lorenz gauge invariance," or covariance under transformations $x'^\mu = a_\nu{}^\mu(x) x^\nu + b^\mu(x)$, which was motivated by Utiyama (19) and developed by Kibble (20).

The covariant derivative may also be written as (21):

$$[\partial'_\mu - \frac{1}{2} a^k{}_j(x') \{\partial'_\mu a^{lj}(x')\} S_{kl}] \psi(x'), \qquad (13)$$

where $S_{kl} = \frac{1}{2} \gamma_k \gamma_l$. We *define* the quantities

$$A^{kl}{}_\mu(x') \equiv - a^k{}_j(x') \partial'_\mu a^{lj}(x') \qquad (14)$$

and notice that because $a^k{}_j(x) a_l{}^j(x) = \delta^k{}_l$ one has the relation

$$A^{kl}{}_\mu(x) = - A^{lk}{}_\mu(x) \qquad (15)$$

These quantities are identified with the "local affine connection" of the bimetric vierbein formalism.

The introduced local affine connection variables $A^{kl}{}_\mu(x)$, enter as "potential" terms in the *covariant derivatives* of spinors:

$$\Psi(x)_{;\mu} = \Psi(x)_{,\mu} + \frac{1}{4} A^{kl}{}_\mu(x)\, \gamma_k \gamma_l\, \Psi(x), \tag{16}$$

$$\bar{\psi}(x)_{;\mu} = \bar{\psi}(x)_{,\mu} + \frac{1}{4} A^{kl}{}_\mu(x)\, \bar{\psi}(x)\, \gamma_k \gamma_l$$

In our subsequent development we will use the standard tensor convention that a comma followed by an index stands for ordinary differentiation with respect to the coordinate denoted by that index and that a semicolon followed by an index similarly denotes *covariant differentiation*.

The local affine connection is also used when one takes the covariant derivatives of quantities referred to the local coordinates:

$$A^k(x)_{;\mu} = A^k(x)_{,\mu} + A^{kl}{}_\mu(x) A_l(x) \tag{17}$$
$$A_k(x)_{;\mu} = A_k(x)_{,\mu} - A^l{}_{k\mu}(x) A_l(x)$$

For the covariant derivatives of global quantities, one would expect a *global affine connection* $\Gamma^\rho{}_{\mu\nu}(x)$, similar to that of Riemannian geometry to enter. Guided by the fact that in Riemannian geometry the covariant derivative of the metric tensor vanishes - i. e. $g^{\mu\nu}(x)_{;\rho} = 0$, we define a global affine connection by requiring that the covariant derivative of the bimetric vierbein must vanish.

$$h^{k\nu}(x)_{;\mu} \equiv 0 \equiv h^{k\nu}(x)_{,\mu} + A^{kl}{}_\mu(x) h_l{}^\nu(x) + \Gamma^\nu{}_{\rho\mu}(x) h^{k\rho}(x), \tag{18}$$

In the absence of matter and S-field interactions, the $\Gamma^\nu{}_{\rho\mu}(x)$ defined here are identical with the Christoffel symbols of General Relativity for the free gravitational field.

The Lagrangian density for the Dirac field can be obtained by replacing all of the derivatives in the usual Dirac Lagrangian by covariant derivatives in the bimetric framework, and by using the vierbein variables $h_j{}^\nu(x)$. (This, of course, is a consequence of our requirement that the theory be globally covariant in the bimetric sense previously described). Thus, by substituting the covariant form of the derivatives from Equation 16, into the usual, Special Relativistic, Dirac Lagrangian:

$$L_{D(sr)} \bar{\psi} \psi + i/2 [\bar{\psi} \gamma^\mu \psi_{,\mu} - \bar{\psi}_{,\mu} \gamma^\mu \psi], \tag{19}$$

We get the generally covariant form, or a Lagrangian dependant on both the S-field and the gravitational variables:

$$L_D = m \bar{\psi} \psi + i/2 [\bar{\psi} \gamma^\mu \psi_{,\mu} - \bar{\psi}_{,\mu} \gamma^\mu \psi] + i/8 \bar{\psi} \gamma^\mu A^{kl}{}_\mu \gamma_k \gamma_l \psi - i/8 A^{kl}{}_\mu \bar{\psi} \gamma_k \gamma_l \gamma^\mu \psi \tag{20}$$

Or because of the antisymmetry and the commutation properties of the γ's,

$$L_D = m\bar{\psi}\psi + i/2[\bar{\psi}\gamma^\mu\psi_{,\mu} - \bar{\psi}_{,\mu}\gamma^\mu\psi] + i/8 A^{kl}{}_\mu \bar{\psi}(\gamma^\mu\gamma_k\gamma_l + \gamma_k\gamma_l\gamma^\mu)\psi \qquad (21)$$

Note that the last term is just the "bimetric local affine connection" coupled to the spin density of the Dirac field.

Using the explicit form of the Dirac action in interaction with the S-field, in the total action of Equation 11, and performing the variation with respect to the Dirac operators $\bar{\psi}(x)$ and $\psi(x)$, leads to the Euler-Lagrange equations:

$$h[m\bar{\psi} - i/2\bar{\psi}_{,\mu}\gamma^\mu + i/4 A_{kl\mu}\bar{\psi}\gamma^\mu\gamma^k\gamma^l] - i/2(h\bar{\psi}\gamma^\mu)_{,\mu} = 0 \qquad (22)$$
$$h[m\psi + i/2\gamma^\mu\psi_{,\mu} + i/4 A_{kl\mu}\gamma^\mu\gamma^k\gamma^l\psi] + i/2(h\gamma^\mu\psi)_{,\mu} = 0$$

or

$$[m\bar{\psi} - i\bar{\psi}_{,\mu}\gamma^\mu] + i/4 A_{kl\mu}\bar{\psi}\gamma^\mu\gamma^k\gamma^l - i/(2h)\bar{\psi}h_{,\mu}h_l^\mu\gamma^l = 0 \qquad (23)$$
$$[m\psi + i\gamma^\mu\psi_{,\mu}] + i/4 A_{kl\mu}\gamma^\mu\gamma^k\gamma^l\psi + i/(2h)h_{,\mu}h_l^\mu\gamma^l\psi = 0$$

where because of the antisymmetry of the $A_{kl\mu}$ and the commutation properties of the $\gamma$'s the integers corresponding to *n, k,* and *l* must all be different. These equations differ from the generally covariant form of the Dirac equations used in the literature by the last term (arising due to the density nature of the Lagrangian), which does not appear if one just writes covariant derivatives in the place of the ordinary derivatives in the usual Dirac equations. These equations of motion imply, just as in the special-relativistic case, that the Dirac Lagrangian density vanishes. This can be seen by multiplying the first part of Equation 23, on the right by $\psi$ and the second on the left by $\bar{\psi}$ and adding the resulting equations to get:

$$2L_D = 2m\bar{\psi}\psi + i[\bar{\psi}\gamma^\mu\psi_{,\mu} - \bar{\psi}_{,\mu}\gamma^\mu\psi] + i/2 A_{kl\mu}\bar{\psi}(\gamma^\mu\gamma^k\gamma^l)\psi = 0 \qquad (24)$$

We are now in the position to derive the equations for the interacting S-field. Analogously with the vierbein-affine connection formulation of the Dirac field interacting with the gravitational field (15). One can introduce a bimetric "contracted curvature tensor density," obtained, in the usual way by considering the commutator of two covariant derivatives. Consider Equation 16:

$$\Psi(x)_{;\mu} = \Psi(x)_{,\mu} + \frac{1}{4}A^{kl}{}_\mu(x)\gamma_k\gamma_l\Psi(x),$$

One can note that $\Psi(x)_{;\mu}$ is a vector-spinor transforming as a first-rank tensor under general coordinate transformations, and as a spinor under local spin transformations; therefore,

$$\Psi_{;\mu;\nu} = (\Psi_{;\mu})_{,\nu} - \Gamma^{\rho}{}_{\mu\nu}(\Psi_{;\rho}) + \frac{1}{4}A^{kl}{}_{\nu}\gamma_k\gamma_l(\Psi_{;\mu}) \tag{25}$$

Substituting the covariant derivatives from Equation 17, gives:

$$\Psi_{;\mu;\nu} = \Psi_{,\mu\nu} + \frac{1}{4}A^{kl}{}_{\nu}\gamma_k\gamma_l\Psi_{,\mu} + \frac{1}{4}A^{kl}{}_{\mu}\gamma_k\gamma_l\Psi_{,\nu} - \Gamma^{\rho}{}_{\mu\nu}\Psi_{;\rho} + \frac{1}{16}A^{kl}{}_{\nu}\gamma_k\gamma_l A^{mn}{}_{\mu}\gamma_m\gamma_n\Psi + \frac{1}{4}A^{kl}{}_{\mu,\nu}\gamma_k\gamma_l\Psi \tag{26}$$

Now if we similarly form

$$\Psi_{;\nu;\mu} = \Psi_{,\nu\mu} + \frac{1}{4}A^{kl}{}_{\mu}\gamma_k\gamma_l\Psi_{,\nu} + \frac{1}{4}A^{kl}{}_{\nu}\gamma_k\gamma_l\Psi_{,\mu} - \Gamma^{\rho}{}_{\nu\mu}\Psi_{;\rho} + \frac{1}{16}A^{kl}{}_{\mu}\gamma_k\gamma_l A^{mn}{}_{\nu}\gamma_m\gamma_n\Psi + \frac{1}{4}A^{kl}{}_{\nu,\mu}\gamma_k\gamma_l\Psi \tag{27}$$

and subtract Equation 24 from Equation 23, the first three terms cancel, and we get

$$\Psi_{;\mu;\nu} - \Psi_{;\nu;\mu} = \frac{1}{4}(A^{kl}{}_{\mu,\nu} - A^{kl}{}_{\nu,\mu})\gamma_k\gamma_l\Psi - (\Gamma^{\rho}{}_{\mu\nu} - \Gamma^{\rho}{}_{\nu\mu})\Psi_{;\rho}$$
$$+ \frac{1}{16}(A^{kl}{}_{\nu}A^{mn}{}_{\mu} - A^{kl}{}_{\mu}A^{mn}{}_{\nu})\gamma_k\gamma_l\gamma_m\gamma_n\Psi \tag{28}$$

Consider the last term of the above expression for the case when all of the Latin indices are different; then $\gamma_k\gamma_l\gamma_m\gamma_n$ is completely antisymmetric, and the whole term vanishes. For the case when two of the $\gamma$'s have the same index, and after collecting terms and changing the dummy index from $n$ to $l$, Equation 28, can be put in the form:

$$\Psi_{;\mu;\nu} - \Psi_{;\nu;\mu} = \frac{1}{4}R^{kl}{}_{\mu\nu}\gamma_k\gamma_n\Psi - (\Gamma^{\rho}{}_{\mu\nu} - \Gamma^{\rho}{}_{\nu\mu})\Psi_{;\rho} \tag{29}$$

where $R^{kl}{}_{\mu\nu}$ is now a "bimetric curvature tensor" defined by:

$$R^{kl}{}_{\mu\nu} \equiv (A^{kl}{}_{\mu,\nu} - A^{kl}{}_{\nu,\mu}) + (A^{km}{}_{\mu}A^{l}{}_{m\nu} - A^{km}{}_{\nu}A^{l}{}_{m\mu}) \tag{30}$$

These equations can be used to construct the explicit Lagrangian density of the S-field in the presence of gravitation analogously to the usual Riemannian case as:

$$L_{IB} = h[h_l{}^{\mu}(x)h_k{}^{\nu}R^{kl}{}_{\mu\nu}] \equiv hR \tag{31}$$

Using a Palatini type of variation, in which the quantities. $h_k{}^{\mu}$ and $A^{kl}{}_{\mu}$ are considered as independent variables and are varied independently, gives two sets of equations.

Variation of the total action, using $L_{IB}$ from Equation 31, and $L_D$ from Equation 21, with respect to the $A$'s, or,

$$\frac{\partial L_{IB}}{\partial A_\mu^{kl}} - \partial_\sigma\left(\frac{\partial L_{IB}}{\partial A_{\mu,\sigma}^{kl}}\right) - \left\langle \left|\frac{\partial L_D}{\partial A_\mu^{kl}} - \partial_\sigma\left(\frac{\partial L_D}{\partial A_{\mu,\sigma}^{kl}}\right)\right| \right\rangle = 0 \tag{32}$$

gives:

$$(hh_k^\alpha h_l^\mu - hh_k^\mu h_l^\alpha)_{,\alpha} - (hh_m^\alpha h_l^\mu - hh_m^\mu h_l^\alpha)A_{k\alpha}^m - (hh_k^\mu h_m^\alpha - hh_k^\alpha h_m^\mu)A_{l\alpha}^m =$$
$$ih/8\, \bar{\psi}\, (\gamma^\mu \gamma_k \gamma_l - \gamma_l \gamma_k \gamma^\mu)\psi \tag{33}$$

The right hand side of Equation 33, can be identified as matter spin density,

$$ih/8\, \bar{\psi}\, (\gamma^\mu \gamma_k \gamma_l - \gamma_l \gamma_k \gamma^\mu)\psi \equiv h S^\mu_{kl}. \tag{34}$$

The left hand side of Equation 33, is similar in form to the Christoffel connection, but now contains the S-field variables interacting with the antisymmetric matter spin density, thus the Christoffel symbols, $\Gamma$, will have an antisymmetric part. The form of Equation 33, implies that the S-field through its bimetric connection, *acts as an information source* of the Dirac field spin density, as we would expect for a field that would impart entanglement related information.

The influence of the S-field can be thought of as a "force" arising from an S-field "potential." Similarly to the case of gravitation when the gravitational "force" appears when one projects the geodesics into Euclidean space, the S-field "force" manifests itself when one projects the geodesics in bimetric space-time onto Euclidean space associated with $\eta^{ij}$, (16).

Variation of the total action, with respect to the $h$'s, gives:

$$\frac{\partial L_{IB}}{\partial h_l^\mu} - \partial_\sigma\left(\frac{\partial L_{IB}}{\partial h_{l,\sigma}^\mu}\right) - \left\langle \left|\frac{\partial L_D}{\partial h_l^\mu} - \partial_\sigma\left(\frac{\partial L_D}{\partial h_{l,\sigma}^\mu}\right)\right| \right\rangle = 0 \tag{35}$$

or,

$$h(h_\mu^l R - 2R_\mu^l) = -\left\langle \left|\frac{\partial L_D}{\partial h_l^\mu} - \partial_\sigma\left(\frac{\partial L_D}{\partial h_{l,\sigma}^\mu}\right)\right| \right\rangle \tag{36}$$

The right-hand side of Equation 36, can be identified with the expectation value of the stress-energy tensor operator for the Dirac field (this would hold for any quantized field in question) taken in the state of the system, i. e.

$$h(h_\mu^l R - 2R_\mu^l) = -\left\langle \left|\mathbf{T}_\mu^l\right| \right\rangle \equiv T_\mu^l \tag{37}$$

Explicitly, this tensor for the Dirac field is:

$$T^l{}_\mu = +i/2[\bar\psi \gamma^l \psi_{,\mu} - \bar\psi_{,\mu} \gamma^l \psi] + (i/4)A_{kj\mu}\bar\psi \gamma^l \gamma^k \gamma^j \psi \tag{38}$$

Others have similarly argued that non-linear fields (particularly gravitation) should be regarded as classical fields determined by quantized sources (22), (23), (24).

It is readily seen that in the absence of matter-S-field interactions the only contribution to the bimetric is from the free gravitational field and variation with respect to $h_k{}^\mu(x)$, gives equations of motion equivalent to the usual Einstein equations for the free-field:

$$\frac{\partial L_{IB}}{\partial h_l^\mu} - \partial_\sigma\left(\frac{\partial L_{IB}}{\partial h_l^\mu{}_{,\sigma}}\right) = 0 \tag{39}$$

Because $\dfrac{\partial h}{\partial h_l^\mu} = -h\, h^l{}_\mu$, Equation 38, is just:

$$h h^l{}_\mu R - 2 h h_k{}^\nu R^{kl}{}_{\mu\nu} = 0 \,. \tag{40}$$

which in turn implies:

$$R^l{}_\mu = 0 \tag{41}$$

One can derive conservation laws for the general form of the "bimetric" stress-energy tensor of Equation 37. The stress-energy tensor of the Dirac field in the presence of the S-field and gravitation no longer satisfies a conservation law of the form:

$$T^\mu{}_{\nu;\mu} = 0 \tag{42}$$

A globally covariant (in bimetric space-time) conservation law for the Dirac field in the presence of the S-field and gravitation becomes:

$$T^\mu{}_{\nu;\mu} + (S^\alpha{}_{\nu\mu} - \tfrac{1}{2}\delta^\mu{}_\nu S^\lambda{}_{\lambda\mu} - \tfrac{1}{2}\delta^\mu{}_\nu S_\nu{}^\lambda{}_\lambda)T^\mu{}_\alpha + R^{kl}{}_{\mu\nu} S^\mu{}_{kl} = 0 \tag{43}$$

Using the definitions for $R$ from Equations 30 and 31, we can define geometrical variables in the bimetric space-time,

$$B^\mu{}_\nu \equiv R^\mu{}_\nu - \tfrac{1}{2}\delta^\mu{}_\nu R, \tag{44}$$

in terms of which one can write the above conservation law for the bimetric stress-energy tensor in manifestly covariant form:

$$B^\mu{}_{\nu;\mu} = 0. \tag{45}$$

Analogous to the usual Einstein-Cartan theory, we can also define the current, and the four-momentum vectors in the bimetric space time.

The current vector is defined as:

$$J^\mu \equiv i\left(\frac{\partial L_D}{\partial \psi_{;\mu}}\psi - \frac{\partial L_D}{\partial \bar\psi_{;\mu}}\bar\psi\right) = h\bar\psi\gamma^\mu\psi \qquad (46)$$

and by virtue of the antisymmetry of the $\Gamma$, satisfies the conservation law:

$$J^\mu_{;\mu} = 0 = \partial_\mu(h\bar\psi\gamma^\mu\psi) \qquad (47)$$

The four-momentum can be formed from Equation 44, can be used to form the analogue of the total four-momentum vector:

$$P_\mu = \int h h_j^\nu B_\mu^j d\sigma_\nu \qquad (48)$$

Where the integral is over a Tomonaga-Schwinger type, space-like hypersurface (25).

**4. Conclusions**

In summary, one can structure a theory in terms of a *classical* bimetric space-time, where the gravitational field, is seen to be formed by the expectation values of the *operator* stress-energy tensor of the quantized material fields. The superluminal, S-field appears to provide the mediating *information* effecting the spin states through its vierbein variables effecting the matter spin density spinor.

The assumptions as to the c-number nature of the S-field, were partly motivated by the expectation that such a field is a more fundamental nonquantized field as implied in John Bell's quote: *"Could the 'non-superluminal signaling' of 'local' quantum field theory be regarded as an adequate formulation of the fundamental causal structure of physical theory? I do not think so…There are many cases in practice where a field can be considered, to be classical and 'external' to the quantum system… Where are truly 'external' fields to be found? Perhaps at the interface between the brain and the mind?"*(1). In the above quote, Bell is stating a conclusion similar to ours. He also anticipates the existence of a 'truly external' classical field. He uses the words "classical and 'external'," because he expects a new kind of field, and not just a classical 'approximation' of the current quantum fields. In this spirit, we postulated that both gravitation and S-fields should be considered as classical fields with c-number variables. From this point of view, the material "quantized" fields, operate against the background of classical S-field and gravitational metrics or more generally, a bimetric, affine, space-time.

In measuring the usual observables of the material fields, only averages over small volumes of space-time have "physical" meaning, since material fields propagate with limited speed and thus require finite times for their forces to manifest given effects on classical test bodies used in current physics (26). Information passed with the help of material fields of quantum theory and their probabilistic measurements through "reduction of the state vector" in Hilbert space, will continue be limited by the speed of light.

However, assuming an ontological existence of the postulated S-field and focusing on coherent phase interactions, may lead to new insights as to superluminal information transfer. It is likely, however, that our notion of information and its measurements will need to be modified to include an extension of the density matrix formulations of entropy.

In conclusion, others modeled quantum entanglement, and the apparent superluminal quantum information effects, by assuming that it was the gravitational or electromagnetic fields which mediated such information exchange (9), (11), (14), or they introduced a preferred frame of reference (16), (27). This requires one to experimentally observe Lorenz symmetry breaking for the gravitational and the electromagnetic fields, which to date has not been found (28), or to work in the Bohm formulation of quantum field theory, which is more complex. The approach taken in this paper, accepts the current proven aspects of relativistic quantum field theories, while providing a formulation that could causally explain the quantum information phenomena, and lead to further research and insights.